\RequirePackage{fix-cm}
\documentclass[smallextended]{svjour3}       
\smartqed  
\usepackage{graphicx}
\begin{document}

\title{Tunable scattering cancellation of light using anisotropic cylindrical cavities}

\titlerunning{Scattering cancellation using anisotropic cavities}        

\author{Carlos D\'{\i}az-Avi\~n\'o$^1$ \and Mahin Naserpour$^{1}$ \and Carlos J. Zapata-Rodr\'{\i}guez$^{1,*}$}

\authorrunning{Carlos D\'{\i}az-Avi\~n\'o et al.}

\institute{
	$^1$ Department of Optics and Optometry and Vision Science, University of Valencia, Dr. Moliner 50, Burjassot 46100, Spain \\
	$^*$ Corresponding author: \email{carlos.zapata@uv.es}
}

\date{Received: 13 April 2016}

\maketitle

\begin{abstract}
Engineered core-shell cylinders are good candidates for applications in invisibility and cloaking.
In particular, hyperbolic nanotubes demonstrate tunable ultra-low scattering cross section in the visible spectral range.
In this work we investigate the limits of validity of the condition for invisibility, which was shown to rely on reaching an epsilon near zero in one of the components of the effective permittivity tensor of the anisotropic metamaterial cavity.
For incident light polarized perpendicularly to the scatterer axis, critical deviations are found in low-birefringent arrangements and also with high-index cores.
We demonstrate that the ability of anisotropic metallodielectric nanocavities to dramatically reduce the scattered light is associated with a multiple Fano-resonance phenomenon.
We extensively explore such resonant effect to identify tunable windows of invisibility. 
\end{abstract}

\section{Introduction}

Cloaking and invisibility are optical techniques with considerable advances recently due to the advent of metamaterials.
Designs for cloaking where a shadow region prevents the light-matter interaction with a tailored target placed therein are largely based on transformation optics \cite{Leonhardt06,Pendry06,Cai07} providing extensive theoretical studies and physical analysis without drawing on numerical simulations \cite{Forouzeshfard15,Yu15}.
On the other hand, invisibility relies on the scattering cancellation of a given object by using for instance a metallic coating and even complex nanostructured coverings \cite{Alu10,Ni10}; the negative polarizability of the carpet layer might severely drop the scattering cross section of the particle making it undetectable \cite{Alu05}.
The first experimental realization was performed in the microwave spectral range by using an array of metallic fins, immersed in a high-permittivity environment, thus creating a metamaterial cloaking shell \cite{Edwards09}.
Of particular interest results the inclusion of shells made of materials with a permittivity near zero in the spectral range of interest, which may lead to a significant drop of the scattering spectrum and, in addition, create a shielding effect in the bounded space \cite{Filonov12}.

The development of multilayered plasmonic coatings and metasurfaces integrates the current state of the art in the engineering of devices for applications in invisibility and cloaking \cite{Alu08,Tricarico09,Chen12}. 
A recent proposal proving a critical reduction of the scattering cross section consists of alternating metallic and dielectric coatings which are arranged in a periodic radial distribution, thus shaping an anisotropic-nanostructured hyperbolic shell \cite{Kim15}. 
For the purpose of simplifying the description of the stratified metamaterial and the application of the analytical Lorenz-Mie scattering method \cite{Bussey75,Bohren98}, a long-wavelength approximation is used thus enabling an adequate interpretation of the resulting spectra under some circumstances \cite{Diaz16}.
A key issue is that the permittivity tensor describing the nanotube reaches near zero values of one of its components in the vicinities of the invisibility regime.

In this study we analyze in detail the validity of such an approach and we extend their results to achieve a higher tunability degree concerning the invisibility spectral windows.
For that purpose, we first evaluate the scattering cross section for different configurations of a nanostructured infinitely-long shell and for different core and environment media in order to verify the existence of a minimum in the epsilon-near-zero regime.
For the sake of generality, we assume a Drude model for the characterization of the material with negative permittivity, and we employ the effective medium theory to describe the form anisotropy of the hollow cylinder.
Furthermore, we establish a matrix-transfer formulation with applications in multilayered radially-anisotropic media, which presents some similarities to developments previously implemented in stratified plane metamaterials \cite{Yeh77}.  
We demonstrate that in the narrow band with epsilon near zero cylinders, the invisibility of the particle becomes a reality provided that the core and environment medium are not polarizable, and that the Drude medium composing the shell has a moderate and low filling factor.
Additional higher-energy bands of scattering reduction are found for large particles with applications in invisibility.

\section{The Lorenz-Mie scattering coefficients}

\begin{figure*}[tb]
	\includegraphics[width=\linewidth]{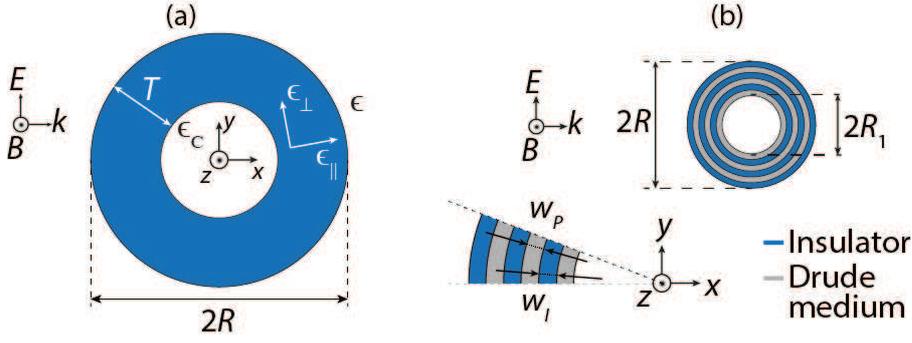}
	\caption{
		(a) Illustration of the anisotropic infinitely-long cavity.
		(b) Coaxial multilayered metamaterial establishing a radial-form birefringence.
	}
	\label{fig01}
\end{figure*}

We consider a cylindrical shell formed by a radially anisotropic nanostructure.
The optical arrangement is illustrated in Fig.~\ref{fig01}(a).
The permittivities along the optic axis (OA), $\epsilon_\parallel$, which is radially directed, and perpendicular to the OA, $\epsilon_\perp$, constitute the components of the permittivity tensor $\underline{\epsilon} = \epsilon_\parallel \hat{r} \hat{r} + \epsilon_\perp \hat{\theta} \hat{\theta} + \epsilon_\perp \hat{z} \hat{z}$.
In principle, these permittivities may be complex valued, thus taking into account losses in the metamaterial, and their real part may take a positive, a negative, and even a near zero value.  
In our numerical simulations we considered an anisotropic tube of utmost radius $R$ with a shell thickness given by $T < R$.
In this study we examined tubes with a core material with permittivity $\epsilon_C$ and immersed in an environment medium of permittivity $\epsilon$.

To estimate analytically the scattering efficiency of the anisotropic nanotube, we followed the Lorenz-Mie scattering method given for instance in Refs.~\cite{Chen12b} and \cite{Chen13}.
First we assumed that the nanotube is illuminated by a TE$^z$-polarized plane wave propagating along the $x$ axis, as illustrated in Fig.~\ref{fig01}(a).
The magnetic field of the incident plane wave may be set as 
\begin{equation}
\mathbf{B}^i = \hat{z} B_0 \exp \left( i k x \right) ,
\end{equation}
where $B_0$ is a constant amplitude, $k = k_0 \sqrt{\epsilon}$ and $k_0 = \omega/c$ is the wavenumber in the vacuum.
In this case, the scattered magnetic field in the environment medium, $r > R$, may be set as \cite{Bohren98}
\begin{equation}
\mathbf{B}^s = \hat{z} B_0 \sum_{n = - \infty}^{+\infty} a_n i^n H_n^{(1)} \left(k r \right) \exp \left( i n \phi \right) ,
\label{eq06}
\end{equation}
where $r$ and $\phi$ are the radial and azimuthal cylindrical coordinates, respectively, and $H_n^{(1)}$ is the Hankel function of the first kind and order $n$.
The total magnetic field in the environment medium is simply $\mathbf{B}^{(tot)} = \mathbf{B}^i + \mathbf{B}^s$.

In the anisotropic shell (medium 2), $R_1 < r < R$, where $R_1 = R - T$, the magnetic field may be set as \cite{Chen12b} 
\begin{equation}
\mathbf{B}^{(2)} = \hat{z} B_0 \sum_{n = - \infty}^{+\infty} i^n \left[ b_n J_{n'} \left(k_2 r \right) + c_n Y_{n'} \left( k_2 r \right) \right] \exp \left( i n \phi \right) ,
\end{equation}
where $J_{n'}$ and $Y_{n'}$ are the Bessel functions of the first and second kind, respectively, both of the order $n'$ given by
\begin{equation}
n' = \sqrt{\frac{\epsilon_\perp}{\epsilon_\parallel}} n ,
\end{equation}
and the wavenumber $k_2 = k_0 \sqrt{\epsilon_\perp}$.
Finally, the magnetic field in the core of the anisotropic tube, which corresponds to the medium 1 ($r < R_1$), is expressed as
\begin{equation}
\mathbf{B}^{(1)} = \hat{z} B_0 \sum_{n = - \infty}^{+\infty} i^n d_n J_n \left(k_1 r \right) \exp \left( i n \phi \right) ,
\end{equation}
where the wavenumber $k_1 = k_0 \sqrt{\epsilon_C}$.

The Lorenz-Mie scattering coefficients $a_n$, $b_n$, $c_n$, and $d_n$, are determined by means of the proper boundary conditions, that is, continuity of the $z$-component of the magnetic field and the $\phi$-component of the electric field, established at the environment-anisotropic medium interface given at $r = R$ and at the core-anisotropic medium interface set at $r = R_1$.
In particular, the boundary conditions applied at $r = R_1$ may be set in matrix form as
\begin{equation}
D_{n,1}^Y \left( R_1 \right) \cdot \left[ \begin{array}{c} d_n \\ 0 \end{array} \right] 
=
D_{n',2}^Y \left( R_1 \right) \cdot \left[ \begin{array}{c} b_n \\ c_n \end{array} \right]  ,
\end{equation}
where the matrix
\begin{equation}
D_{n,m}^Y \left( x \right) = \left[ \begin{array}{cc} J_n \left( k_m x \right) & Y_n \left( k_m x \right) \\ Z_m J'_n \left( k_m x \right) & Z_m Y'_n \left( k_m x \right)   \end{array} \right]
\end{equation}
is given in terms of the reduced impedance $Z_1 = 1 / \sqrt{\epsilon_C}$ and $Z_2 = 1 / \sqrt{\epsilon_\perp}$ for the media $m = 1$ and $m = 2$, respectively.
Here the prime appearing in $J'_n \left( \alpha \right)$ and $Y'_n \left( \alpha \right)$ denotes derivative with respect to the variable $\alpha$.
By applying the boundary conditions at $r = R$ we may write
\begin{equation}
D_{n}^H \left( R \right) \cdot \left[ \begin{array}{c} 1 \\ a_n \end{array} \right]
=
D_{n',2}^Y \left( R \right) \cdot \left[ \begin{array}{c} b_n \\ c_n \end{array} \right]  ,
\end{equation}
where 
\begin{equation}
D_{n}^H \left( R \right) = \left[ \begin{array}{cc} J_n \left( k R \right) & H_n^{(1)} \left( k R \right) \\ Z J'_n \left( k R \right) &  Z {H^{(1)}}'_n \left( k R \right)   \end{array} \right] ,
\end{equation}
where $Z = 1 / \sqrt{\epsilon}$.

Finally, we may estimate the fields in the core space and outside the nanotube without calculating the fields in the anisotropic medium by means of the following matrix equation:
\begin{equation}
\left[ \begin{array}{c} 1 \\ a_n \end{array} \right] =
M_n \cdot \left[ \begin{array}{c} d_n \\ 0 \end{array} \right]  ,
\end{equation}
where the matrix
\begin{eqnarray}
M_{n} &=& \left[ \begin{array}{cc} M_{n,11} & M_{n,12} \\ M_{n,21} &  M_{n,22}   \end{array} \right] \\
&=& \left[ D_{n}^H \left( R \right) \right]^{-1} \cdot D_{n,2}^Y \left( R \right) \cdot \left[ D_{n,2}^Y \left( R_1 \right) \right]^{-1} \cdot D_{n,1}^Y \left( R_1 \right) . \nonumber
\end{eqnarray}
By using this transfer matrix formalism, it is possible to evaluate analytically the scattering coefficients 
\begin{equation}
a_n = \frac{M_{n,21}}{M_{n,11}} ,
\end{equation}
which provide an exact estimation of the scattering efficiency as
\begin{equation}
Q_s = \frac{2}{k R} \sum_{n = - \infty}^{+\infty} |a_n|^2 .
\label{eq07}
\end{equation}
The nanotube resonances are determined by the poles of the coefficients $a_n$, that is for the zeros of $M_{n,11}$.
On the other hand, the invisibility condition is established provided that the scattering coefficients $a_n$ (or alternatively $M_{n,21}$) arrive simultaneously to a value near zero.

\section{Metamaterial with form birefringence}

From a practical point of view, a radially anisotropic medium may be established by the proper form birefringence of a metamaterial composed of concentric multilayers as illustrated in Fig.~\ref{fig01}(b) \cite{Torrent09,Kettunen15}.
Here, two materials with permittivity of opposite sign were used for the stratified medium in order to substantially increase the form birefringence \cite{Kidwai12}.
A plasmonic nanofilm of width $w_\mathrm{P}$ is set by the side of an insulator layer of width $w_\mathrm{I}$, thus forming the unit cell of a periodic distribution along the radial coordinate.
In our numerical simulations we considered a nanostructured tube of utmost radius $R$ and composed of a number of subwavelength layers giving a total shell thickness $T$.
The permittivity of the plasmonic material, given by
\begin{equation}
\epsilon_\mathrm{P}(\omega) = 1 - \frac{\omega_p^2}{\omega^2 + i \omega \gamma} ,
\label{eq01}
\end{equation}
follows the Drude model within the spectral range of interest \cite{Maier07}. 
In the previous equation, $\omega_p$ represents the plasma frequency and $\gamma$ stands for the damping rate.
The real part of $\epsilon_\mathrm{P}(\omega)$ is negative provided that $\omega^2 < \omega_p^2 - \gamma^2$, the latter condition approaching $\omega < \omega_p$ for a low-loss plasmonic material.
For simplicity we will consider a nondispersive permittivity $\epsilon_\mathrm{I}$ for the insulator in the frequency range under study.

For our structured metamaterial with a deeply-subwavelength period, the medium may be considered as a uniaxial crystal within the limits of the long-wavelength approximation \cite{Yeh88,Elser07}.
The optic axis of the metamaterial is set along the direction of periodicity; in our particular case, the OA is oriented along the radial axis.
The effective anisotropic medium is then optically characterized by a local permittivity tensor $\underline{\epsilon}$ of component \cite{Yeh88}
\begin{equation}
\epsilon_\parallel (\omega) = \frac{ \epsilon_\mathrm{I} \epsilon_\mathrm{P} (\omega)}{f \epsilon_\mathrm{I} + (1 - f) \epsilon_\mathrm{P}(\omega)} ,
\label{eq03}
\end{equation}
along the OA, and 
\begin{equation}
\epsilon_\perp (\omega) = f \epsilon_\mathrm{P} (\omega) + (1 - f) \epsilon_\mathrm{I} ,
\label{eq04}
\end{equation}
in the perpendicular direction.
In the previous equations, the filling factor of the Drude medium in the metamaterial is given by
\begin{equation}
f = \frac{w_\mathrm{P}}{w_\mathrm{P} + w_\mathrm{I}} ,
\label{eq05}
\end{equation}
which represents the unique geometrical parameter determining the effective permittivities $\epsilon_\parallel$ and $\epsilon_\perp$ of the metamaterial, regardless the internal one-dimensional distribution of the materials involved in the unit cell.

\begin{figure*}[tb]
	\includegraphics[width=\linewidth]{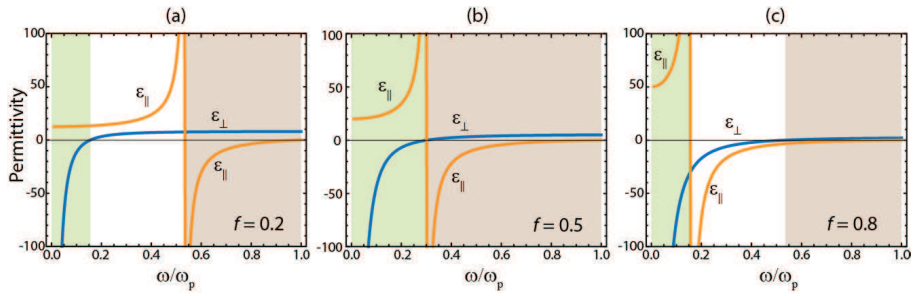}
	\caption{Real part of the components $\epsilon_\parallel$ (orange line) and $\epsilon_\perp$ (blue line) of the permittivity tensor for the metamaterial composed of a Drude medium with $\gamma = \omega_p / 100$ and an insulator of permittivity $\epsilon_\mathrm{I} = 10$, assuming different filling factors: (a) $f = 0.2$, (b) $f = 0.5$, and (c) $f = 0.8$.
		The shaded regions denote spectral bands where the metamaterial exhibits a hyperbolic dispersion of the Type I (shaded in mauve) and the Type II (shaded in green).}
	\label{fig02}
\end{figure*}

In Fig.~\ref{fig02} we represent the real values of $\epsilon_\parallel$ and $\epsilon_\perp$ of a metamaterial composed of a Drude medium and an insulator of permittivity $\epsilon_\mathrm{I} = 10$, evaluated within a range of frequencies near $\omega_p$ and below, considering different values of the filling factor $f$.
We observe that the real part of $\epsilon_\perp$ is near zero around a frequency
\begin{equation}
\omega_\mathrm{zero} = \frac{ \sqrt{ \left( \omega_p^2 - \gamma^2 \right) f - \gamma^2 ( 1 - f ) \epsilon_\mathrm{I}} } { \sqrt{ \epsilon_\mathrm{I} - f ( \epsilon_\mathrm{I} - 1 )}} ,
\label{eq18}
\end{equation}
whereas the real part of $\epsilon_\parallel$ is near a pole around 
\begin{equation}
\omega_\mathrm{pole} = \omega_p \frac{ \sqrt{ 1 - f }}{ \sqrt{ f ( \epsilon_\mathrm{I} - 1 ) + 1}} ,
\label{eq19}
\end{equation}
the latter being valid when $\gamma \ll \omega_p$.
The hyperbolic regime is determined by the condition $\mathrm{Re} (\epsilon_\parallel) \mathrm{Re} (\epsilon_\perp) < 0$, indicated as shaded regions in Fig.~\ref{fig02}.
The choice $\mathrm{Re} (\epsilon_\perp) > 0$ corresponds to the so-called Type I hyperbolic metamaterials, whereas the choice $\mathrm{Re} (\epsilon_\perp) < 0$ denotes a Type II hyperbolic medium \cite{Ferrari15}. 
Note that when $f = 1/2$, corresponding to the case that the plasmonic and insulator layers have the same width, both a zero of $\epsilon_\perp$ and a pole of $\epsilon_\parallel$ occurs simultaneously at a frequency $\omega_p / \sqrt{1 + \epsilon_\mathrm{I}}$, in addition happening when $\mathrm{Re} (\epsilon_\mathrm{P}) = - \epsilon_\mathrm{I}$.
In this case, the hyperbolic regime spans the whole spectrum below the plasma frequency.
Let us point out that the extraordinary dispersion features of hyperbolic and epsilon-near-zero metamaterials have inspired us in a plethora of novel applications such as subwavelength imaging \cite{Zapata11e,Zapata12a}, surface-wave engineering \cite{Miret10,Zapata13b}, and double refraction \cite{Zapata14b,Diaz16}, to mention a few.

\section{Results and discussion}

\begin{figure}[tb]
	\centering
	\includegraphics[width=.6\linewidth]{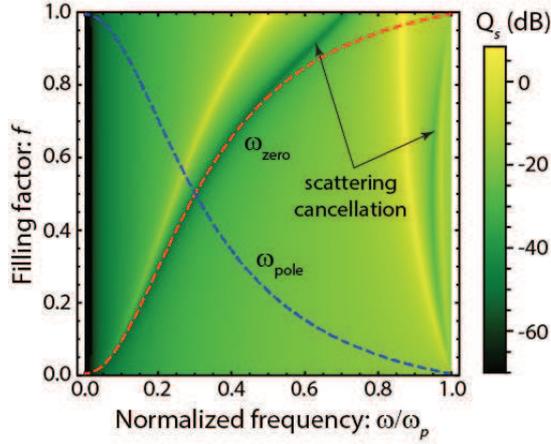}
	\caption{
		Scattering efficiency $Q_s$, expressed in dB, of an anisotropic cavity of equal inmost radius and thickness $R_1 = T = k_p^{-1} / 20$, immersed in air ($\epsilon = \epsilon_1 = 1$).
		The filling factor of the Drude medium varies from 0 to 1, considering a fixed damping rate $\gamma = \omega_p / 100$.
		The calculations are performed for TE$^z$-polarized incident light.
		The red dashed line corresponds to frequencies $\omega_\mathrm{zero}$ which are solutions to the equation~(\ref{eq18}), whereas the blue dashed line denotes frequencies $\omega_\mathrm{pole}$ given in Eq.~(\ref{eq19}). 
	}
	\label{fig03}
\end{figure}

In Fig.~\ref{fig03} we represent the scattering efficiency $Q_s$ in dB as derived from Eq.~(\ref{eq07}) and calculated by means of the transfer matrix method described above, that is assuming a TE$^z$-polarized incident light.
The scattering spectrum is evaluated in terms of the $\epsilon$-negative material filling factor $f$, assuming a low-loss Drude medium with $\gamma = \omega_p / 100$.
The anisotropic cavity is immersed in air, where $\epsilon = \epsilon_C = 1$, and it is small enough ($R_1 = T = k_p^{-1} / 20$, where $k_p = \omega_p / c$) to excite only the dipole term $n = 1$ of the series given in Eq.~(\ref{eq07}).
We observe two resonances corresponding to the symmetric and antisymmetric coupling between the surface charges associated with the cavity and surface polaritons at $r = R_1$ and $r = R$ \cite{Zhu08,Nickelson12}.
By the side of each resonance peak, a minimum in scattering at a slightly higher frequency is found.
This effect has been also found in solid high-index cylinders and wires, which is attributed to the existence of Fano resonances \cite{Rybin15}.
Importantly, the set of \emph{fundamental} frequencies where the scattering cancellation is shown at the lowest energy, is set virtually over the curve $\omega_\mathrm{zero}$ given in Eq.~(\ref{eq18}), coinciding with a metamaterial permittivity $\epsilon_\perp$ near zero.
We point out that such outcome has been previously reported by Kim \emph{et al} in Ref.~\cite{Kim15}.
Since $\epsilon_\perp$ spans the whole spectrum below $\omega_p$ by balancing the composition of the Drude medium, in principle we might tune the invisibility frequency by simply changing the value of $f$ from 0 to 1. 

However, such a procedure cannot be followed for invisibility bands near the plasma frequency, even considering values of the filling factor $f$ close to unity.
Kim's approach is not valid for filling factors $f$ close to unity, where the tubular particle behaves like a low-birefringent plasmonic cavity.
In this case, the scattering cancellation associated with the fundamental peak is observed at an intermediate frequency between the symmetric and antisymmetric resonances found below $\omega_p$; in particular, the invisibility frequency is found at $\omega = 0.707 \omega_p$ in the limit $f \to 1$.
On the other hand $\omega_\mathrm{zero}$ approaches the plasma frequency of the Drude medium in such a limit.

Importantly, proper conditions for invisibility are found at higher energies provided that the filling factor $f$ takes moderate values.
Specifically, such scattering cancellation may be observed in the vicinities of the plasma frequency $\omega_p$. 
From Fig.~\ref{fig02} it is clear that, in order to achieve a scattering cancellation near $\omega_p$, a second minimum of $Q_s$ which is located close to the plasma frequency might be also exploited.
For instance, at $f = 1/2$ we find the first minimum in the scattering efficiency, $Q_s = 3.07 \times 10^{-7}$, at the fundamental frequency $\omega = 0.3 \omega_p$, and a secondary minimum $Q_s = 5.11 \times 10^{-5}$ located at a frequency $\omega = 0.945 \omega_p$.
Here we conclude that invisibility at the lowest energy provides a better performance than the reduction in scattering found near the plasma frequency.

\begin{figure}[tb]
	\centering
	\includegraphics[width=.6\linewidth]{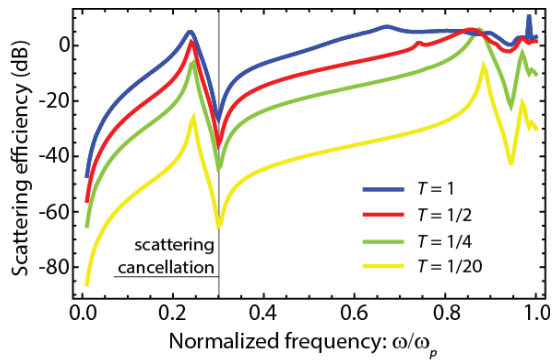}
	\caption{
		Scattering efficiency of metamaterial cavities immersed in air, where the aspect ratio $T / R = 1 / 2$ is conserved. 
		The tube thickness $T$ is expressed in units of $k_p^{-1}$.
		The anisotropic metamaterial is again characterized by a filling factor of the Drude medium set as $f=1/2$, where the damping rate is $\gamma = \omega_p / 100$.
	}
	\label{fig04}
\end{figure}

In Fig.~\ref{fig04} we show the scattering efficiency of anisotropic cavities of a higher size.
In particular, we set a filling factor $f = 1/2$, and compare the case analyzed in Fig.~\ref{fig03} with those where the inmost radius and the tube thickness take the values $R_1 = T = k_p^{-1} / 4$, $R_1 = T = k_p^{-1} / 2$ and $R_1 = T = k_p^{-1}$.
Regarding the spectral position of the fundamental resonance and the lowest-energy cancellation of scattering, all curves show essentially the same behavior.
For instance, the invisibility frequency is found at $\omega = 0.3 \omega_p$ in all cases with valley efficiencies given by $Q_s = 3.7 \times 10^{-5}$, $2.7 \times 10^{-4}$, and $2.3 \times 10^{-3}$, as long as the tubular particle increases in size.
As a consequence, the invisibility window is invariant upon the cavity diameter, provided that the modulation is performed in the subwavelength scale.
As expected, although the scattering efficiency reaches a minimum at the mentioned frequency, such efficiency grows approximately one order of magnitude when the tube doubles in size.

On the other hand, the situation changes dramatically when analyzing the scattering spectrum near the secondary minimum.
The pattern of the scattering efficiency is maintained unaltered for the smallest cavities, except for a scaling factor.
However, the onset of additional resonances at increasing sizes (in the order of the current wavelength) governs the contour of the spectrum in the vicinity of the plasma frequency.
Since the minima of the high-order Fano resonances, which in nature present a lower depth than the fundamental resonance, overlap with the peaks of the neighbors thus vanishing the invisibility effect.

\begin{figure}[tb]
	\centering
	\includegraphics[width=.6\linewidth]{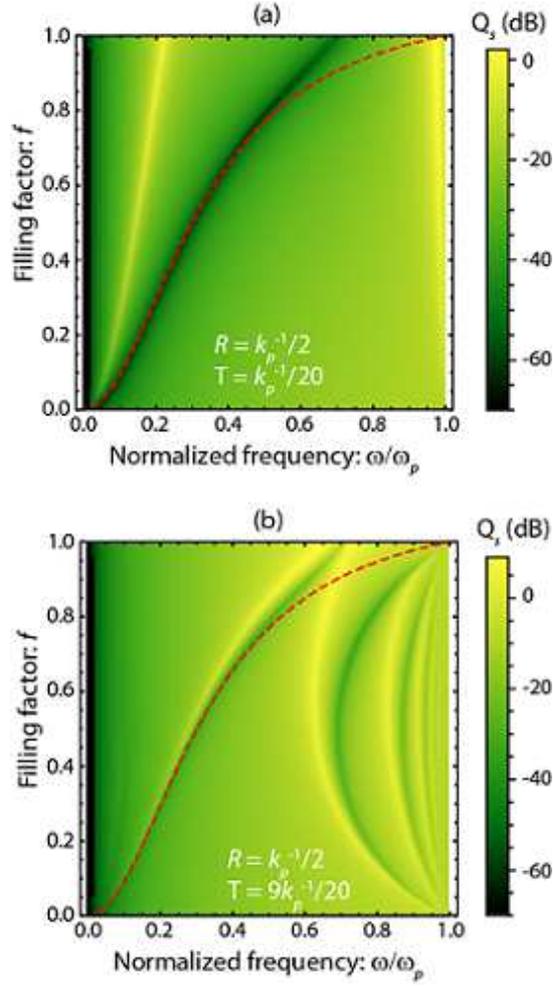}
	\caption{
		Scattering spectrum of a tubular cavity immersed in air ($\epsilon = \epsilon_1 = 1$) with utmost radius $R = k_p^{-1} / 2$, when the thickness of the shell is (a) $T = 0.1 R$, and (b) $T = 0.9 R$.
		The red dashed line indicates the frequency $\omega_\mathrm{zero}$ given in Eq.~(\ref{eq18}).
	}
	\label{fig05}
\end{figure}

The spectrum of the scattering efficiency with respect to the filling factor $f$ of the metamaterial, for cylinders of radius $R = k_p^{-1} / 2$ and different geometrical configurations, i.e. various values of the aspect ratio $T/R$, is depicted in Fig.~\ref{fig05}.
First we analyzed the response of an anisotropic cavity with a very narrow shell.
In Fig.~\ref{fig05}(a) we represent the scattering efficiency when $T = R/10$.
The position of the two polaritonic resonances are clearly displaced with respect to the cases shown above.
In particular, the secondary peak is located in the very-close neighborhood of the plasma frequency.
The latter has a dramatic consequence: the secondary frequency associated with scattering cancellation drops out of sight, limiting the tunability of the invisibility effect.
Notably, the main invisibility frequency remains practically unaltered, approaching $\omega_\mathrm{zero}$ at low and moderate filling factors.

Examining the case where $T = 0.9 R$, which suggests a cylinder with a very small concentric hole, we obtained the scattering efficiency depicted in Fig.~\ref{fig05}(b).
The main feature of the derived pattern is the occurrence of multiple peaks and their associated minima in efficiency, indicating an accumulation of excited Fano resonances \cite{Rybin15}.
Nevertheless, the fundamental frequency of invisibility again continues attached to the condition of $\epsilon_\perp$ near zero.

\begin{figure}[tb]
	\centering
	\includegraphics[width=.6\linewidth]{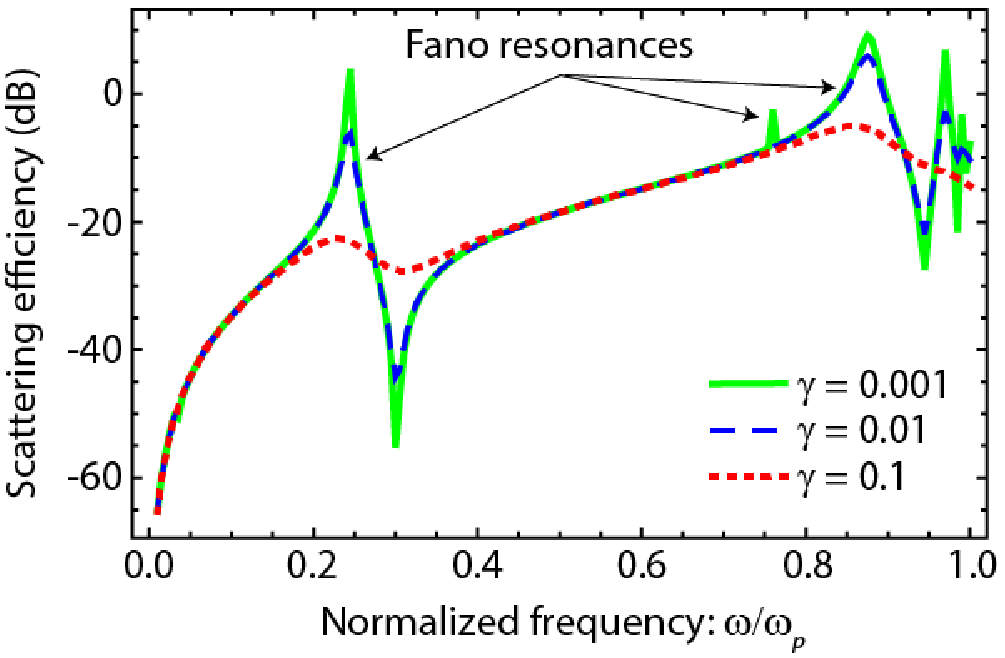}
	\caption{
		Scattering efficiency of a tubular anisotropic metamaterial ($f = 1/2$) of equal radius and thickness, $R = T = k_p^{-1} / 4$, immersed in air.
		The damping rate of the Drude medium varies from $\gamma = 1 / 10$, expressed in units of $\omega_p$, and $\gamma = 1 / 1000$.
	}
	\label{fig06}
\end{figure}

The influence of losses in the Drude medium are analyzed in the following.
In Fig.~\ref{fig06} we represent the scattering efficiency of a hollow metamaterial cylinder immersed in air and with dimensions given by $R = k_p^{-1} / 2$ and $T = R/2$.
Again, the filling factor is $f = 1/2$.
We observe that metamaterials including an ultra low-loss Drude medium, like that assuming $\gamma = \omega_p / 1000$ and analyzed in Fig.~\ref{fig06}, gives a minimum in scattering which represents approximately one order of magnitude lower than that obtained for a damping rate of $\gamma = \omega_p / 100$.
The inverse occurs when examining the peaks corresponding to the fundamental resonance.
As a consequence, the peak-to-valley contrast increases two orders of magnitude for a decrement of only one order of magnitude in the damping rate of the Drude medium.
Regimes for ultra-cancellation of scattering are inevitably related with strong Fano resonances observed in low-loss Drude media.
Also note that additional ultra-narrow peaks emerge, which may be attributed to multipole Mie resonances occurring for high orders $n > 1$.
However, this cannot be applied to achieve an invisibility effect in practice.

On the other hand, considering higher losses of the Drude medium, the minimum and maximum of scattering efficiency associated with the fundamental Fano resonance may reduce their contrast in several orders of magnitude.
This is illustrated in Fig.~\ref{fig06} by evaluating the scattering efficiency of an anisotropic metamaterial including a Drude medium with $\gamma = \omega_p / 10$.
This critical issue is of relevance since the electrons in the Drude medium experience a significant reduction of their effective mean free path in realistic multilayered anisotropic metamaterials; note that electron scattering in ultrathin layers results in the loss of electron phase coherence leading to a higher damping rate \cite{Zhu08}.
Finally, the drop of peak-to-valley contrast may be also observed in higher energy Fano resonances.

\begin{figure*}[tb]
	\includegraphics[width=\linewidth]{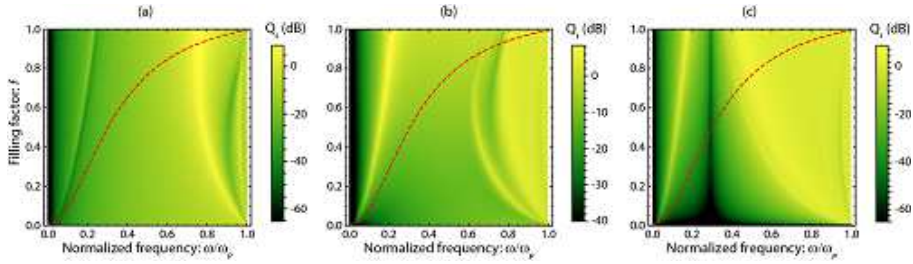}
	\caption{
		Scattering efficiency, expressed in dB, of a cylindrical cavity of radius $R = k_p^{-1} / 2$ and thickness $T = R / 2$, when varying the dielectric constant of the core and environment medium: (a) $\epsilon_C = 10$ and $\epsilon = 1$, (b) $\epsilon_C = 1$ and $\epsilon = 10$, and (c) $\epsilon_C = 10$ and $\epsilon = 10$.
		The red dashed line corresponds to frequencies $\omega_\mathrm{zero}$ which are solutions to the equation~(\ref{eq18}). 
	}
	\label{fig07}
\end{figure*}

Heretofore, the only discrepancy to the rule of $\epsilon_\perp$ near zero, enabling to find the principal Fano resonance with associated scattering cancellation, was found with cavity metamaterials of a high filling factor $f$.
However, that is not certainly unique.
As an illustration we examine the effects of the dielectric constant in the core ($\epsilon_C$) and in the environment medium ($\epsilon$).
In particular, we evaluated the scattering efficiency of a cylindrical cavity of radius $R = k_p^{-1} / 2$ and shell thickness $T = k_p^{-1} / 4$.
Again, a fixed damping rate $\gamma = \omega_p / 100$ is considered.
In Fig.~\ref{fig07}(a) we plot $Q_s$ for cylinders of different filling factors $f$ immersed in air and having a high permittivity core of $\epsilon_C = 10$.
Two resonances associated with the excitation of symmetric and antisymmetric surface polaritons, showing a peak-to-valley spectral pattern, govern again the scattering efficiency of the core-shell cylinder.
The loci of the maxima are clearly shifted with respect to the case of a hollow cavity with  $\epsilon_C = 1$.
Remarkably, the fundamental frequency of invisibility undergoes a substantial departure from $\omega_\mathrm{zero}$ even at low and moderate filling factors.

In the reversed scene, where the environment medium presents a high permittivity, for instance $\epsilon = 10$ as shown in Fig.~\ref{fig07}(b), and the core is filled with air, the behavior in scattering changes completely.
Firstly, we identify the fundamental Fano resonance with an enormous spectral gap between its peak frequency and the first valley frequency, where the peak-to-valley efficiency has a moderate contrast.
When the filling factor $f = 1/2$, the main peak is found at $\omega = 0.153 \omega_p$, whereas the first valley is located at $\omega = 0.707 \omega_p$.
This fact suggests that the reduction of $Q_s$ associated with the fundamental Fano resonance is deleted in practical terms, enabling to observe exclusively minima of higher order resonances in the scattering spectrum.
In this sense, additional resonances covering the spectrum below $\omega_p$ may be found.
We conclude that the effect of invisibility is basically erased from this picture. 

Finally we analyzed the scattering efficiency of the anisotropic cavity for a high dielectric constant of the core and environment medium, simultaneously.
In Fig.~\ref{fig07}(c) we represent $Q_s$ when $\epsilon_C = \epsilon = 10$.
Importantly, the fundamental frequency of invisibility is set practically in the same value for the whole range of filling factors.
For instance, the scattering cancellation is produced at $\omega = 292 \omega_p$ when the filling factor is $f = 0.2$; increasing such a factor to $f = 0.8$, the frequency corresponding to the valley of lowest energy is found at $\omega = 0.294 \omega_p$.
These results demonstrate that the invisibility regime dramatically depends on the permittivities of the core and environment medium, a conclusion that might be in disagreement with previous studies which remarked the importance of the condition given by $\omega_\mathrm{zero}$ \cite{Kim15}.

\section{Conclusions}

In summary, we investigated the scattering efficiency of cylindrical cavities formed by a radially anisotropic nanostructure, when the illumination is carried out by a TE$^z$ plane wave.
A radially anisotropic medium was established by the form birefringence of a metamaterial composed of periodic distribution of subwavelength concentric multilayers.
The Lorenz-Mie scattering coefficients are determined by means of the proper boundary conditions, which we set in terms of a transfer matrix formalism.

For subwavelength cavities, we observe two resonances corresponding to modal symmetric and antisymmetric surface polaritons.
By the side of each resonance peak, a minimum in scattering at a slightly higher frequency is found, which are attributed to Fano shapes.
For scatterers immersed in air, the fundamental frequency where the scattering cancellation is shown at the lowest energy makes the metamaterial permittivity $\epsilon_\perp$ takes values near zero.
Thus, by balancing the composition of the metamaterial, we might tune the invisibility frequency up to the plasma frequency of the constituting Drude medium. 
Unfortunately, such approach is not valid for high filling factors close to unity, where the tubular particle behaves like a low-birefringent plasmonic cavity.
In these cases, the secondary dip in the scattering spectrum might efficiently be used to obtain an invisibility effect.

The onset of additional resonances at increasing cavity sizes, in the order of the current wavelength, governs the contour of the scattering spectrum thus vanishing the invisibility effect.
The main invisibility frequency remains practically unaltered of various values of the aspect ratio $T/R$.
In particular, examining the case which suggests a cylinder with a very small concentric hole, we obtained the occurrence of multiple peaks and their associated minima in efficiency, indicating an accumulation of excited Fano resonances.

Finally, severe discrepancies to the rule of $\epsilon_\perp$ near zero, enabling to find the principal Fano resonance with associated scattering cancellation, were found by modifying the dielectric constant in the core and in the environment medium, even at low and moderate filling factors.
For instance, for a high dielectric constant of the core and environment medium, the fundamental frequency of invisibility is set practically fixed for the whole range of filling factors.
These results demonstrate that the invisibility regime dramatically depends on the permittivities of the core and environment medium.

\begin{acknowledgements}
This work was supported by the Spanish Ministry of Economy and Competitiveness (MINECO) (TEC2014-53727-C2-1-R) 
\end{acknowledgements}

\bibliographystyle{spphys}       

\end{document}